\documentstyle[12pt]{article}
\newcommand{\be}{\begin{equation}}
\newcommand{\ee}{\end{equation}}
\newcommand{\ba}{\begin{eqnarray}}
\newcommand{\ea}{\end{eqnarray}}
\newcommand{\oh}{\displaystyle{\frac{1}{2}}}
\begin{document}
%%%%%%%%%%%%%%%%%%%%%%%%%%%%%%%%%%%%%%%%%%%%%%%%%%%%%%%%%%%%%%%%%
\title          {Fermionic determinant as an overlap
                      between bosonic vacua.}
\author{C.D.\ Fosco$^a$\thanks{CONICET}\\
and\\
F.A.\ Schaposnik$^b$\thanks{Investigador CICBA, Argentina}
\\
{\normalsize\it
$^a$Centro At\'omico Bariloche,
8400 Bariloche, Argentina}\\
{\normalsize\it
$^b$Departamento de F\'\i sica, Universidad Nacional de La Plata}\\
{\normalsize\it
C.C. 67, 1900 La Plata, Argentina}}

\date{~}

\maketitle

\vspace{-4.5 in}
\hfill\vbox{
\hbox{~~}
\hbox{\it Modified version}
\hbox{\today}
}
\vspace{5 in}

%============================================
\begin{abstract}
We find a representation for the determinant of a Dirac
operator in an even number $D= 2 n$ of Euclidean dimensions as an
overlap between two different vacua, each one corresponding to a
bosonic theory with a quadratic action in $2 n + 1$ dimensions,
with identical kinetic terms, but differing in their mass terms.
This resembles the overlap representation of a fermionic
determinant (although {\em bosonic\/} fields are used here).
This representation may find applications to lattice field theory,
as an alternative to
other bosonized representations for Dirac determinants already
proposed.
\end{abstract}
\newpage
%============================================
Based on an earlier idea of Kaplan~\cite{kap}, the overlap
formalism~\cite{nar,overl} has been proposed as a way to define
fermionic chiral determinants.  Its lattice implementation seems
to overcome the kinematical constraints imposed by the
Nielsen-Ninomiya theorem~\cite{kars}. It may thus provide a suitable
framework to study interesting non-perturbative phenomena in models
containing chiral fermions.

In the overlap approach, the determinant of a chiral Dirac operator
in $D=2 d$ dimensions is defined as the overlap, i.e., scalar product,
between the Dirac vacuum states of two auxiliary Hamiltonians acting
on Dirac fermions in $2 d + 1$ dimensions. This method has been recently
extended to the case of Dirac determinants in odd dimensions~\cite{odd1,odd2}.

The overlap between two vacua can be implemented in at least two
different ways: an `operatorial' version, based on building
up the Dirac vacua by occupying all the respective negative energy
states,
and then computing the scalar product; and also in a path integral
approach, which introduces an extra dimension into the game.
The ovelap is  obtained by calculating a path integral amplitude for
a system whose Lagrangian has a mass that depends on the coordinate
labeling points in the extra dimension (in the spirit
of the domain-wall picture).

In this letter we shall provide an alternative definition for
a fermionic determinant as an overlap between two {\em bosonic\/}
vacua. The issue of bosonizing a Dirac operator, namely,
writing a fermionic determinant in $D$ dimensions as a functional
integral over purely bosonic fields\footnote{This is all what is
meant by `bosonization' here. Note that this meaning is quite
different to the one used, for example, in \cite{bos}.} has
recently received atention~\cite{lusch1,slav}.
The approaches \cite{lusch1,slav} share the property of involving
an infinite number of bosonic fields, although for different reasons
than in the overlap approach.
The infinite number of fields manifests itself as an extra dimension
in \cite{slav}, and as a (discrete) infinite number of fields
in the approach of \cite{lusch1}. Besides, they deal not with
the chiral case, but rather with systems involving both chiralities,
i.e., Dirac fermions,
the same case we shall consider in the present work.
This is,  in practice, the interesting
case for lattice $QCD$. In treatments that avoid the quenched
approximation, a bosonic representation for the quark
determinant (which is real) may be very useful indeed. We will, in
this letter, also deal with this case.

~

In our construction, it is convenient to write the vacuum state in the
functional Schr\"odinger representation~\cite{funct} (we shall later on
discuss an operatorial representation). In the Schr\"odinger
representation, the "wavefunction" that describes the vacuum state
for a real scalar field $\varphi$ with a quadratic action is, in the
"coordinate" basis, a vacuum functional $\langle \varphi |\Psi_\Omega
\rangle =
\Psi_\Omega (\varphi)$, of the following kind
\be
\Psi_\Omega (\varphi) \;=\; {\rm det}^{\frac{1}{4}}
(\frac{\Omega}{\pi})\,
\exp (-\frac{1}{2} \varphi \Omega \varphi )
\label{psiom}
\ee
where we are using a shorthand notation (similar to the
one of \cite{zinn}), such that
\be
\varphi \Omega \varphi \;\equiv\; \int d^D x d^D y
\, \varphi (x) \Omega (x,y) \varphi (y)
\label{2}
\ee
and $\Omega (x,y)$ is a real, symmetric, and definite positive kernel.

In the coordinate basis, all the wavefunctions depend on the field
configuration $\varphi (x)$, where $x$ labels the $D$ `spatial'
coordinates in a $D+1$ dimensional spacetime. Thus the scalar
product between two states $\Phi_1$, $\Phi_2$ is defined by
a functional integral
\be
\langle \Phi_1 | \Phi_2 \rangle \;=\; \int {\cal D} \varphi \;
\Phi_1^* (\varphi) \, \Phi_2 (\varphi)
\label{scalar}
\ee
where the integration measure in (\ref{scalar})
is formally defined as ${\cal D} \varphi \equiv \prod_x d \varphi_x$
with $x$ in the $D$ dimensional space.

The functional determinant factor in (\ref{psiom}) is introduced
in order to normalize this wavefunction
\be
\langle \Psi_\Omega | \Psi_\Omega \rangle \;=\; 1 \;.
\ee

This vacuum state can be thought of as the ground state for a
(second quantized) Hamiltonian ${\cal H}(\pi,\varphi)$
\be
{\cal H} (\pi , \varphi) \;=\;
\oh \left[ \int d^Dx \,\pi^2 (x) + \int d^Dx \int d^Dy \,
\varphi (x) \Omega^2 (x,y) \varphi (y) \right]
\ee
with $\pi (x) = {\delta}/{i \delta \varphi (x)}$
the momentum operator, and $\varphi$ acting  multiplicatively.
 Following notation (\ref{2}), we write
$\Omega^2 (x,y) \,=\, \int d^Dz \Omega (x,z) \Omega (z,y)$.
As an example,  one has for a free real scalar field
$\Omega^2 (x,y) = (-\nabla_x^2 + m^2)\delta (x-y)$.

Let us now consider the overlap between two vacuum states,
$\Psi_{\Omega_1}(\varphi)$, $\Psi_{\Omega_2}(\varphi)$,
corresponding to two Hamiltonians, differing just in their
(commuting) kernels $\Omega_1$, $\Omega_2$. Performing a
functional Gaussian integration, and rearranging terms,
the overlap
$\langle \Psi_{\Omega_1} | \Psi_{\Omega_2}\rangle$ yields
$$
\langle \Psi_{\Omega_1} | \Psi_{\Omega_2} \rangle \;=\;
{\rm det}^{\frac{1}{4}} (\frac{\Omega_1 \Omega_2}{\pi^2}) \;
{\rm det}^{-\frac{1}{2}}(\frac{\Omega_1 + \Omega_2}{2 \pi})
$$
\be
=\; {\rm det}^{-\frac{1}{4}} \left[ \frac{1}{4}
(\frac{\Omega_1}{\Omega_2}+\frac{\Omega_2}{\Omega_1} + 2)
\right] \;.
\label{overr}
\ee
Or, by defining ${\cal O} \equiv \frac{\Omega_1}{\Omega_2}$,
\be
\langle \Psi_{\Omega_1} | \Psi_{\Omega_2} \rangle \;=\;
{\rm det}^{\frac{1}{4}} [ f ({\cal O}) ]
\label{overf}
\ee
where
\be
f (x) \;=\; \frac{x}{(1 + x)^2} \;,
\label{deff}
\ee
and we have ignored (and so will do in what follows) irrelevant
constant factors. Determinants are supposed to be adequately regularized

so that all usual properties (like $\det(AB) = \det A . \det B)$ )
hold.
The symmetry     $f(x) = f(\frac{1}{x})$ is a consequence
of the symmetry of the
overlap (\ref{overr})
 under $ 1 \leftrightarrow 2$.

The generalization to the case
of a complex scalar field, endowed with a complex hermitean definite
positive covariance $\Omega$, is straightforward. One obtains, instead
of (\ref{overf}), the result
\be
\langle \Psi_{\Omega_1} | \Psi_{\Omega_2} \rangle \;=\;
{\rm det}^{\frac{1}{2}} [ f ({\cal O}) ] \;.
\label{overc}
\ee
This of course generalizes to the case of $N_f$ complex flavours. The
results implies
just to a change of
the exponent of the determinant in the form
(\ref{overc}) to ${\rm det}^{\frac{N_f}{2}}[f({\cal O})]$.

As usual in similar approaches, we shall assume that the
operators $\Omega_1$, $\Omega_2$ have been regularized
(on the lattice, for example), what makes them bounded. We
shall also assume there is a non-vanishing gap in their spectra.
Then their ratio ${\cal O}$ is also regularized. Moreover, we
will let it depend on a mass parameter, say $M$, such that
\be
M \to \infty \; \Rightarrow \; ||{\cal O}(M)|| \; << 1 \;
\ee
then
\be
\langle \Psi_{\Omega_1} | \Psi_{\Omega_2} \rangle \;\sim\;
{\rm det}^{\frac{1}{2}} [ \frac{\Omega_1}{\Omega_2} ] \;\;
{\rm when} \;\; M \to \infty
\;.
\label{limf}
\ee

To deal with the specific example of Dirac operators, we will
consider the case of $\Omega_1$ and $\Omega_2$ such that
\be
\Omega_1^2 \;=\; - {\not \!\! D}^2 + m^2 \;\;,\;\;
\Omega_2^2 \;=\; - {\not \!\! D}^2 + M^2
\ee
where $\not \!\! D \,=\, \gamma_\mu D_\mu$, $D_\mu = \partial_\mu
+ A_\mu$, and $A_\mu$ is an (Abelian or non-Abelian) gauge
connection. To apply the result (\ref{limf}) to the present
case, we need to assume that $\not \!\! D$ is regularized
(in the lattice, say). Then, keeping the regulator cutoff finite,
and letting $M >> m$, $M >> ||\not \!\! D||$, we have, for
the case of $N_f$ flavours,
\be
\langle \Psi_{\Omega_1} | \Psi_{\Omega_2} \rangle
\; \sim \;
{\rm det}^{\frac{N_f}{4}}
[\frac{- {\not \!\! D}^2 + m^2 }{- {\not \!\! D}^2 + M^2} ]
\; \sim \;
{\rm det}^{\frac{N_f}{4}}
[\frac{- {\not \!\! D}^2 + m^2 }{M^2} ] \;.
\label{app1}
\ee
Note that the regulating cutoff must be kept finite when taking
this large $M$ limit, and indeed, $M$ should be in fact bigger
than the cutoff.
Of course, if we normalize by dividing (\ref{app1}) by
the same object evaluated for zero external field ($A=0$),
we obtain
\be
\frac{\langle\Psi_{\Omega_1}|\Psi_{\Omega_2}\rangle}{\langle
\Psi_{\Omega_1}|\Psi_{\Omega_2}\rangle|_{A=0}}
\; \sim \;
{\rm det}^{\frac{N_f}{4}}
[\frac{- {\not\!\! D}^2 + m^2 }{ - {\not \!\partial}^2 + m^2} ] \;,
\label{app2}
\ee
Of course, the number $N_f$ of bosonic
flavours in the overlap must be adjusted, depending on the actual
number of {\em fermionic\/} flavours $N_F$ considered. It is
evident that we can only consider an even $N_f$, and moreover
$N_f = 2 N_F$. Then,we shall write our main result as
\be
{\rm det}^{\frac{N_F}{2}}
[\frac{- {\not\!\! D}^2 + m^2 }{ - {\not \!\partial}^2 + m^2} ]
\sim
\frac{\langle\Psi_{\Omega_1}|\Psi_{\Omega_2}\rangle}{\langle
\Psi_{\Omega_1}|\Psi_{\Omega_2}\rangle|_{A=0}}
\label{main}
\ee
We have then been able to write the
Dirac operator determinant for a $D$-dimensional
theory of Dirac fermions  with
$N_F$ flavour in terms of the ratio of two overlaps
between vacuum states in a $D+1$ dimensional
bosonic theory. Note that since we are working with a Dirac
operator in $D=2n$ dimensions, one has the identity
$\det (-{\not\!\! D}^2 + m^2) = \det({\not\!\! D} + m)^2$. Then,
our overlap formula gives in fact the value for
$|det({\not\!\! D} + m)|$. We see that our approach,
as Slavnov's and Luscher's, only works in even dimensions. 
The reason is that only in even dimensions is the Dirac 
determinant real. The extension of our method to odd dimensions 
would imply the necessity of defining
 vacuum functionals for non-Hermitian Hamiltonians. 

There is a striking similarity between this case,
for a finite $M$ (see equation (\ref{overf})), and what the
standard overlap yields for the modulus of a chiral determinant:
$|\langle A + | A - \rangle| = {\rm det}^{\frac{1}{2}}
\left(\frac{C^\dagger C}{C^\dagger C + M^2}\right)$, where
$C$ is the chiral Dirac operator~\cite{ft}.

For the concrete example of the Dirac operator in $D=2$ dimensions,
we may even interpret the ratio ${\cal O}$ as a Pauli-Villars
regularization for the operator
../$\Omega_1 \,=\, \sqrt{-{\not \!\! D}^2 +m^2}$, since just one
regulator field suffices (the contribution of
this regulator is of course $\Omega_2$). We still assume a lattice
regularization. Then the overlap for this case is doubly regulated,
when one takes the large-$M$ limit (in particular, $M >> 1/a$,
where $a$ is the lattice spacing), one is removing the Pauli-Villars
regulator, and at the same time approaching the determinant. The latter
is still regulated since $a$ is kept finite.

We will also give an operatorial construction of the overlap
(\ref{overr}) between the two vacuum states, which avoids the
use of the Schr\"odinger representation. We need to
introduce suitable creation and annihilation operators.
As the Hamiltonian is quadratic, we shall first consider the
case of one simple harmonic oscillator, and then generalize to
the case of interest, which contains an infinite collection of
decoupled harmonic modes.
For a harmonic oscillator, the analogous of (\ref{overr}) would
be to evaluate the scalar product between to vacua, each one
corresponding to a Hamiltonian with a given frequency $\omega$.
Namely, we need to evaluate (using operatorial methods) the
object $\langle \Psi_{\omega_1} | \Psi_{\omega_2} \rangle$
where $|\Psi_{\omega_1}\rangle$ and  $|\Psi_{\omega_2}\rangle$
are the ground states of $H_1=\frac{1}{2}(p^2+\omega_1^2q^2)$
and $H_2=\frac{1}{2}(p^2+\omega_2^2q^2)$, respectively. It is
possible to show that, if we define the corresponding two sets
of creation and annihilation operators $a_i$, $a_i^\dagger$,
$i=1,2$, they will be related by the
Bogoliubov transformation. This transformation is exactly of the
kind that appears when defining "squeezed states" \cite{umez}:
\ba
a_1&=&\cosh\theta\,a_2\;+\;\sinh\theta\,a_2^\dagger
\;=\; U_\theta\, a_2 U_\theta^\dagger \nonumber\\
a_1^\dagger &=&\sinh\theta\,a_2\;+\;\cosh\theta\,a_2^\dagger
\;=\;U_\theta\, a_2^\dagger U_\theta^\dagger
\ea
where
\ba
\cosh \theta &=& \oh \left(\sqrt{\frac{\omega_1}{\omega_2}}
                          +\sqrt{\frac{\omega_2}{\omega_1}}\right)
\nonumber\\
\sinh \theta &=& \oh \left(\sqrt{\frac{\omega_1}{\omega_2}}
                          -\sqrt{\frac{\omega_2}{\omega_1}}\right)
\ea
and
\be
U_\theta \;=\; \exp [ \frac{1}{2} \theta (a_2^2 - (a_2^\dagger)^2)] \;.
\ee
>From this it follows that we can write one vacuum in terms of the
other as $\Psi_{\omega_1}\rangle \,=\,
U_\theta |\Psi_{\omega_2}\rangle$. It is evident that
$|\Psi_{\omega_1}\rangle$, when represented in the Hilbert space
built with $a_2^{\dagger}$ acting on $|\Psi_{\omega_2}\rangle$,
will be a linear combination of states containing an even number
of excitations. After some algebra, this can be put more explicitly
as follows~\cite{umez}
\be
|\Psi_{\omega_1}\rangle \;=\; e^{-\oh \ln \cosh \theta} \,
 e^{-\oh (a^\dagger)^2 \tanh \theta} \, |\Psi_{\omega_2}\rangle \;.
\label{vr}
\ee
>From (\ref{vr}), we obtain for the overlap
\be
\langle \Psi_{\omega_1} | \Psi_{\omega_2} \rangle \;=\;
 e^{-\oh \ln \cosh \theta} \;=\;
[ 2 f(\frac{\omega_1}{\omega_2}) ]^{\frac{1}{4}}
\label{overq}
\ee
with $f$ as defined in (\ref{deff}). This is clearly the
equivalent of (\ref{overf}) for the case of a $0+1$ field
theory. But the generalization to the $D+1$ case is
trivial because the Hamiltonians are quadratic, and the
system is brought to an infinite collection of
uncoupled harmonic oscillators by using operators that create
or destroy particles occupying states that are eigenmodes
of the kernels $\Omega_1^2$ and $\Omega_2^2$ (they are Hermitian,
and differ by an operator proportional
to the identity). These, of course, are the usual plane waves when
$\Omega^2 (x,y) = (-\nabla_x^2 + m^2)\delta (x-y)$.

Defining thus the eigenmodes $g_k (x)$ by
\be
\int d^Dy \;\Omega_1^2(x,y) g_k (y) =
\lambda_k^2 g_k (x) \;\;,\;\;
\int d^Dy \Omega_2^2(x,y) g_k (y) \;=\;
{\lambda'}_k^2 g_k (x)
\ee
we obviously have $\lambda_k^2 - {\lambda'}_k^2 \,=\, M^2 - m^2$.
Then we can apply (\ref{overq}) to each mode, which now has frequency
$\lambda_k^2$ (or ${\lambda'}_k^2$). This yields
\be
\langle \Psi_{\Omega_1} | \Psi_{\Omega_2} \rangle \;=\;
\prod_{\lambda_k}
[ f(\frac{\lambda_k^2}{{\lambda'}_k^2})]^{\frac{1}{4}}
\;=\; {\rm det}^{\frac{1}{4}} [ f ({\cal O}) ]
\ee
in agreement with (\ref{overf}).

We conclude mentioning that our approach shares with the overlap
definition and with the proposals of \cite{slav,lusch1}
the property of involving an infinite number of fields, here
manifested in the use of  a system living in $D+1$ dimensions.
Our method uses bosons rather than fermions, what means that the $D+1$
dimensional theory would be rather unphysical, as in
\cite{lusch1,slav}, in the sense that the action attributed to
the bosons would be quite exotic. It may however, be a useful
technique for lattice simulations, even in the framework of the
Schr\"odinger functional approach, which can be implemented on
the lattice~\cite{lusch2}. This because our representation would allow to include the effect of fermion loops
without having to simulate fermions, a well known stumbling block for
numerical simulations. Of course, one may also attempt to solve for the
vacuum functional analyticaly, for example by variational techniques, and
then calculate the overlap.

\underline{Acknowledgements}: F.A.S. is
partially  suported by CICBA and CONICET,
Argentina and a Commission of the European Communities
contract No:C11*-CT93-0315. C.D.F. is supported by CONICET.
We acknowledge Professors H. Neuberger and  S. Randjbar-Daemi
for comments on a preliminary version of this letter.

\end{document}